\documentclass[twocolumn,nofootinbib,amsmath,amssymb,a4paper]{revtex4}

\usepackage{graphicx}
\usepackage{dcolumn}
\usepackage{amsmath}
\usepackage{bm}

\input{babarsym.tex}

\def\KKKz      {\ensuremath{K^+ K^- K^0}\xspace}

\def\KKK      {\ensuremath{\Kp \Km \Kp}\xspace}

\def\KKKspm      {\ensuremath{ {\Bz}_{\scriptscriptstyle(+-)} }\xspace}
\def\KKKszz      {\ensuremath{ {\Bz}_{\scriptscriptstyle(00)} }\xspace}
\def\KKKll      {\ensuremath{ {\Bz}_{\scriptscriptstyle (L)} }\xspace}

\def\mKK       {\ensuremath{m_{\Kp\Km}}\xspace}

\def\betaeff   {\ensuremath{\beta_{\mathit{eff}}}\xspace}
\def\Acp   {\ensuremath{{A}_{\CP}}\xspace}
\def\cosH      {\ensuremath{\cos \theta_H}\xspace}

\def\BCP      {\ensuremath{\B_{\CP}}} 
\def\Btag      {\ensuremath{\B_{\mathrm{tag}}}} 
\def\qtag      {\ensuremath{q_{\mathrm{tag}}}} 
\def\AorAbar    {\kern 0.18em\optbar{\kern -0.18em {\cal A}}{}\xspace}
\def\forfbar    {\kern 0.18em\optbar{\kern -0.18em f}{}\xspace}
\def\Bflav     {\ensuremath{B_{\rm flav}}\xspace}

\begin{document}

\boldmath
\title{Measurement of angle $\beta$ with time-dependent \CP asymmetry
in $\Bz\to\Kp\Km\Kz$ decays}
\unboldmath

\author{E. Di Marco}
 \email{emanuele.dimarco@roma1.infn.it}
\affiliation{%
Dipartimento di Fisica, Universit\`{a} di Roma ``La Sapienza'', P.le Aldo Moro 2, 00185 Roma, Italy
}%

\begin{abstract}
We present recent results on \CP-violation, and the determination of CKM angle $\beta$, 
with the decay $\Bz\to\Kp\Km\Kz$, with \babar{} and Belle detectors.
\end{abstract}

\maketitle

\section{Introduction}
In the Standard Model (SM) of particle physics, the phase of the
Cabibbo-Kobayashi-Maskawa (CKM) quark-mixing matrix~\cite{bib:C,bib:KM} is the
only source of \CP violation in the quark sector. Due to the
interference between mixing and decay, this phase can be observed in
measurements of time-dependent \CP asymmetries of \Bz
mesons. In the SM, \CP asymmetries in $\b \to s
\bar s s$ decays, such as $\Bz \to \Kp\Km\Kz$, are expected to be nearly
equal to those observed in tree-dominated $\b \to \c \cbar \s$
decays~\cite{bib:s2b}. However, because in the SM the
former are dominated by loop amplitudes, new particles in those loops
potentially introduce new physics at the same order as the SM process.
Within the SM, deviations from the expected \CP asymmetries in $\Bz
\to \KKKz$ decays depend on the Dalitz plot position, but are expected
to be small and positive~\cite{cptheory}. 
In particular, for the decay $\Bz\to\phi\Kz$ they are expected to be 
less than 4\%.
\babar\ extracts the time-dependent \CP-violation parameters by
taking into account different amplitudes and phases across the \Bz and
\Bzb Dalitz plots, while Belle measures it separately for 
$\Bz\to\phi\Kz$ and the rest of $\Kp\Km\Kz$ events, neglecting interference 
between intermediate states.

The analyses presented here are based on 347 (535) million \BB pairs 
collected with the \babar\ (Belle) detector at the SLAC \pep2\ (KEKB)
\epem asymmetric-energy collider.
Data are collected on the \FourS resonance, while a fraction of about 
10\% is collected at approximatively 40 \mev below the \FourS 
resonance, and it is used to study the background arising from 
$\epem \to \qqbar (q=u,d,s,c)$ continuum events.
The \babar{} and Belle detectors are described in detail elsewhere~\cite{bib:nim}.

\section{Event Reconstruction}
Events are fully reconstructed combining tracks and neutral clusters 
in the detector to form $\Bz \to \KKKz$, with a $\Kz$ reconstructed as
$\KS\to\pip\pim$ (\KKKspm) (\babar{} and Belle) and $\KS\to\piz\piz$ (\KKKszz), or $\KL$
(\KKKll) (\babar{} only).
In order to select $B$ candidates we use a set of two kinematic variables:
the beam-energy-substituted mass 
$\mes=\sqrt{(s/2+{\bf p}_i \cdot {\bf p}_B)^2/E_i^2+p^2_B}$
($M_{bc}$ for Belle), 
and the energy difference $\DeltaE=E^*_B-\sqrt{s}/2$. 
Here, $(E_i,{\bf p}_i)$ is the
four-vector of the initial \epem{} system, $\sqrt{s} $
is the center-of-mass energy, ${\bf p}_B$ is the
reconstructed momentum of the \Bz{} candidate, and $E_B^*$ is its
energy calculated in the \epem{} rest frame. 
For signal decays,
the \mes{} distribution peaks near the \Bz{} mass with a
resolution of about $2.5$ \mevcc, and the \DeltaE{}
distribution peaks near zero with a resolution of  $10-50$ \mev,
depending on the final state.
For decays with a \KL{}, \KL momentum is not measured, 
but evaluated by constraining the \Bz mass to the nominal 
value \cite{bib:PDG2006} 
In this case only
\DeltaE is used, and it has a resolution of about 3 \mev.
The main background comes from random combinations of particles produced
in continuum $e^+e^-\to q\bar{q}$.
In the CM frame, these events have a jet-like structure, while \B
decays have a nearly isotropic topology. We parameterize this
difference using several variables, providing additional
discrimination between signal and background. 
Another source of 
background comes from decays of $B$ mesons which mimic the signal.
This background is typically more difficult to suppress.
The contribution of these decays is estimated from Monte Carlo simulations.

For each fully reconstructed \Bz meson (\BCP), we use the remaining
particles in the event to reconstruct the decay vertex of the other \B
meson (\Btag) and identify its flavor \qtag. 
A multivariate tagging algorithm determines the flavor of the \Btag{} meson 
in \babar{} data and classifies
it in one of seven mutually exclusive tagging categories depending on the 
presence of prompt leptons, one or more charged kaons and pions~\cite{bib:tagging}.
The performances of this algorithm are measured with a data sample of fully
reconstructed \Bz decays into flavor eigenstates (\Bflav): 
$\Bz\to D^{(*)-} \pip/\rho^+/a_1^+$.
The effective tagging efficiency is $Q\equiv\sum_c \epsilon^c
(1-2w^c)^2=0.304\pm 0.003$ for \babar{} (similar for Belle), where $\epsilon^c$ ($w^c$) is the
efficiency (mistag probability) for events tagged in category $c$.
For Belle, the tagging algorithm returns \qtag{} and the tag quality $r$, which
varies from $r=0$ for no flavor discrimination to $r=1$ for unambiguous flavor assignment.
Events with $r\leq 0.1$ are discarded for the \CP-asymmetry measurement, 
and the others are sorted into six intervals.
The difference $\deltat \equiv t_{\CP} - t_{\mathrm{tag}}$ of the proper
decay times of the \BCP\ and \Btag\ mesons is calculated from the
measured distance between the reconstructed decay vertices and the
boost ($\beta\gamma = 0.56 (0.465)$ for \babar{} (Belle)) of the \FourS. The error on \deltat,
$\sigma_{\deltat}$, is also estimated for each event.
Events are accepted if the calculated \deltat uncertainty is less than 2.5 ps and 
$\vert\deltat\vert<20$ ps. The fraction of events which satisfy these 
requirements is 95\%.

The \babar{} analysis strategy is to 
perform a maximum likelihood fit to the selected \KKKz events with a likelihood
function $\cal L$, which uses as probability density function (PDF) for each event,
${\cal L} \equiv  {\mathcal P}(\mes, \DeltaE) \cdot {\mathcal P}_{\mathrm{Low}} \cdot 
  {\cal P}_{DP}(\mKK, \cos\theta_H, \deltat, \qtag) \times {\cal R}(\deltat, \sigma_{\deltat})$
where $n_i$ is the yield for each category ($i$ = signal, continuum background, and \BB backgrounds), and
${\mathcal R}$ 
is a \deltat\ resolution function with parameters determined in 
the \Bflav data sample.
${\mathcal P}_{\mathrm{Low}}$ is a supplementary PDF used only in the
fits to the region with $\mKK<1.1$ \gevcc discussed below. It depends on the event shape
variables and, for \KKKll only, the missing momentum of the event.
This PDF accounts for the fact that signal decays have a missing 
momentum consistent with the reconstructed \KL direction, while 
for background events it is more isotropically distributed.
We characterize events on the Dalitz plot
in terms of the invariant $\Kp\Km$ mass, \mKK, 
and the cosine of the helicity angle between the $\Kp$ and the
$\Kz$ in the CM frame of the $\Kp\Km$ system, \cosH.

In the Belle approach the likelihood for the event selection is the same, 
without ${\cal P}_{DP}$.
A loose requirement on the likelihood ratio
${\cal R}_{s/b}\equiv {\cal L}_{sig}/({\cal L}_{sig}+{\cal L}_{bkg})$ is applied, 
and a maximum-likelihood fit to observed \deltat distribution is performed to 
the selected events.

\section{Analysis of Dalitz Plot}
Accounting for the experimental efficiency $\varepsilon$, the Dalitz plot PDF for signal events is
\begin{equation}
{\mathcal P}_{DP} = d\Gamma \cdot \varepsilon(\mKK, \cosH) \cdot | J(\mKK) |,
\end{equation}
where the final term is the Jacobian of the transformation for our
choice of Dalitz plot coordinates.

The time- and flavor-dependent decay rate over the Dalitz plot is
\begin{eqnarray}
d\Gamma \propto  \frac{e^{-|\deltat|/\tau}}{4\tau} &\times&  
        \Big[~ \left | {\cal A} \right |^2 + \left | \bar{ {\cal A} } \right |^2   \\
      && + ~ \eta_{\CP} ~ \qtag  ~2 Im \left (\bar{\cal A} {\cal A}^*  e^{-2 i \beta}\right ) \sin \deltamd \deltat \nonumber \\
        && - ~ q \left (\left | {\cal A} \right |^2 - \left | \bar{ {\cal A} } \right |^2 \right ) \cos\deltamd\deltat 
        ~\Big ], \nonumber
\label{eq::dalitz_plot_rate}
\end{eqnarray}
where $\eta_{\CP} = 1~(-1)$ for \KKKspm, \KKKszz (\KKKll). $\tau$ and \deltamd are the lifetime and mixing frequency of the  \Bz meson, respectively~\cite{bib:PDG2006}. The CKM angle $\beta$ enters through \Bz-\Bzb mixing. Actual world average is $\beta \sim 0.38$~\cite{bib:PDG2006}. 
We define the amplitude $\cal A$ ($\bar{\cal A}$) for \Bz (\Bzb) decay as a sum of isobar amplitudes, 
\begin{equation}            
{\cal A} (\bar{\cal A}) = \sum \limits_r c_r (1 \pm b_r) e^{i (\phi_r \pm \delta_r)} \cdot f(\bar f)_r, \label{eq:A}
\end{equation}
where the parameters $c_r$ and $\phi_r$ are the magnitude and phase of the
amplitude of component $r$, and we allow for different isobar
coefficients for $\Bz$ and $\Bzb$ decays through the asymmetry
parameters $b_r$ and $\delta_r$.  
Our model includes the vector meson $\phi(1020)$. 
We include also decays into intermediate scalar mesons:
$f_0(980)$, $X_0(1550)$, and $\chi_{c0}$.
The angular distribution is constant for scalar decays, 
whereas for vector decays is $Z_r = -4 \vec{q} \cdot \vec{p}$, 
where $\vec{q}$ is the momentum of the resonant daughter, 
and $\vec{p}$ is the momentum of the third particle in the resonance frame. 
We describe the line-shape for the $\phi(1020)$, $X_0(1550)$, and $\chi_{c0}$  
using the relativistic Breit-Wigner function~\cite{Aubert:2006nu}.
For the $\phi(1020)$ and $\chi_{c0}$ parameters, we use average
measurements~\cite{bib:PDG2006}. For the $X_0(1550)$ resonance, we use
parameters from the our analysis of the $\Bp \to \KKK$
decay~\cite{Aubert:2006nu}.
The $f_0(980)$ resonance is described with the coupled-channel (Flatt\'e) function~\cite{Aubert:2006nu},
with the coupling strengths for the $KK$ and $\pi\pi$ channels 
taken as $g_\pi=0.165\pm0.018$~\gevcc, $g_K/g_\pi=4.21 \pm 0.33$, 
and the resonance pole mass $m_r=0.965 \pm 0.010 $~\gevcc~\cite{Ablikim:2004wn}.
In addition to resonant decays, we include three non-resonant amplitudes. 
The existing theoretical models, which consider contributions from contact term or
higher-resonance tails~\cite{Cheng:2002qu,Fajfer:2004cx,Cheng:2005ug} do not 
reproduce well the features observed in data.
Therefore we adopt a phenomenological parameterization~\cite{bib:isospin}
and describe the non-resonant terms as an exponential decay:
\begin{equation}
         {\cal A}_{NR} =   \sum_{i\neq j} c_{ij} e^{i\phi_{ij}} e^{-\alpha m^2_{ij}}
	     \cdot (1+b_{NR}) \cdot  e^{i(\beta+\delta_{NR})} \label{eq:nr}
\end{equation}
and similarly for $\bar{\cal A}_{NR}$.

We compute the \CP-asymmetry parameters for component $r$ from the asymmetries in amplitudes ($b_r$) 
and phases ($\delta_r$) given in Eq.~(\ref{eq:A}).  The rate asymmetry is
\begin{equation}
  {\Acp}_r=\frac{|\bar{\cal A}_r|^2 - |{\cal A}_r|^2}{|\bar{\cal A}_r|^2 + |{\cal A}_r|^2} 
  =\frac{-2b_r}{1+b_r^2},
\label{eq:Acp}
\end{equation}
and ${\betaeff}_r = \beta + \delta_r$ is the phase asymmetry.

The fraction for resonance $r$ is computed 
\begin{equation}
  {\cal F}_r ~=~ \frac{ \int d\cos\theta_H ~d\mKK \cdot |J| \cdot (|{\cal A}_r |^2
    +|\bar{\cal A}_r|^2)    }
  { \int  d\cos\theta_H ~d\mKK \cdot |J| \cdot (|{\cal A} |^2+|\bar{\cal A}|^2)    }.
\end{equation}
The sum of the fractions can be larger than one due to negative interference in the 
scalar sector.

\begin{figure}
  \begin{tabular}{ll}
    \includegraphics[width=0.23\textwidth]{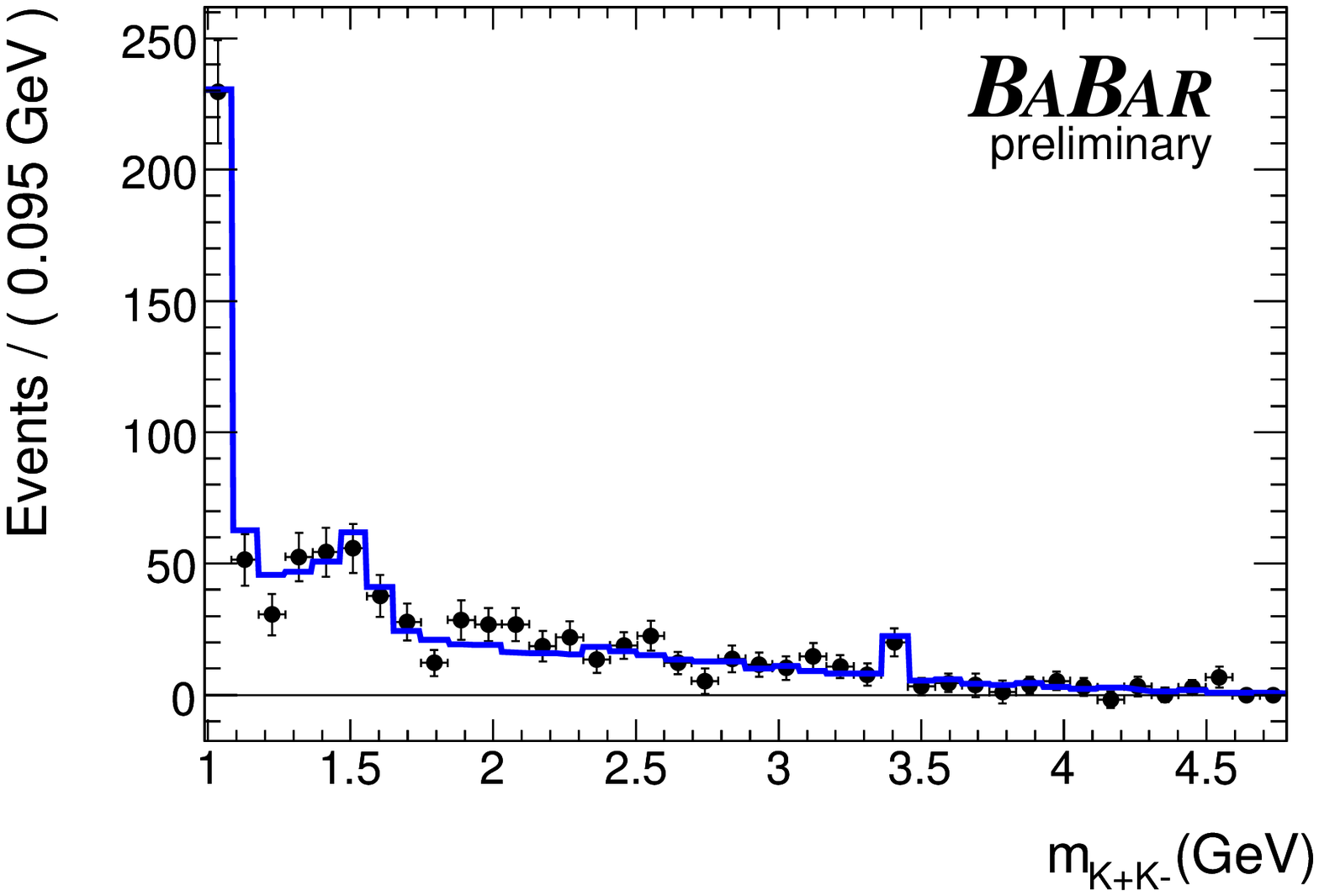} &
    \includegraphics[width=0.23\textwidth]{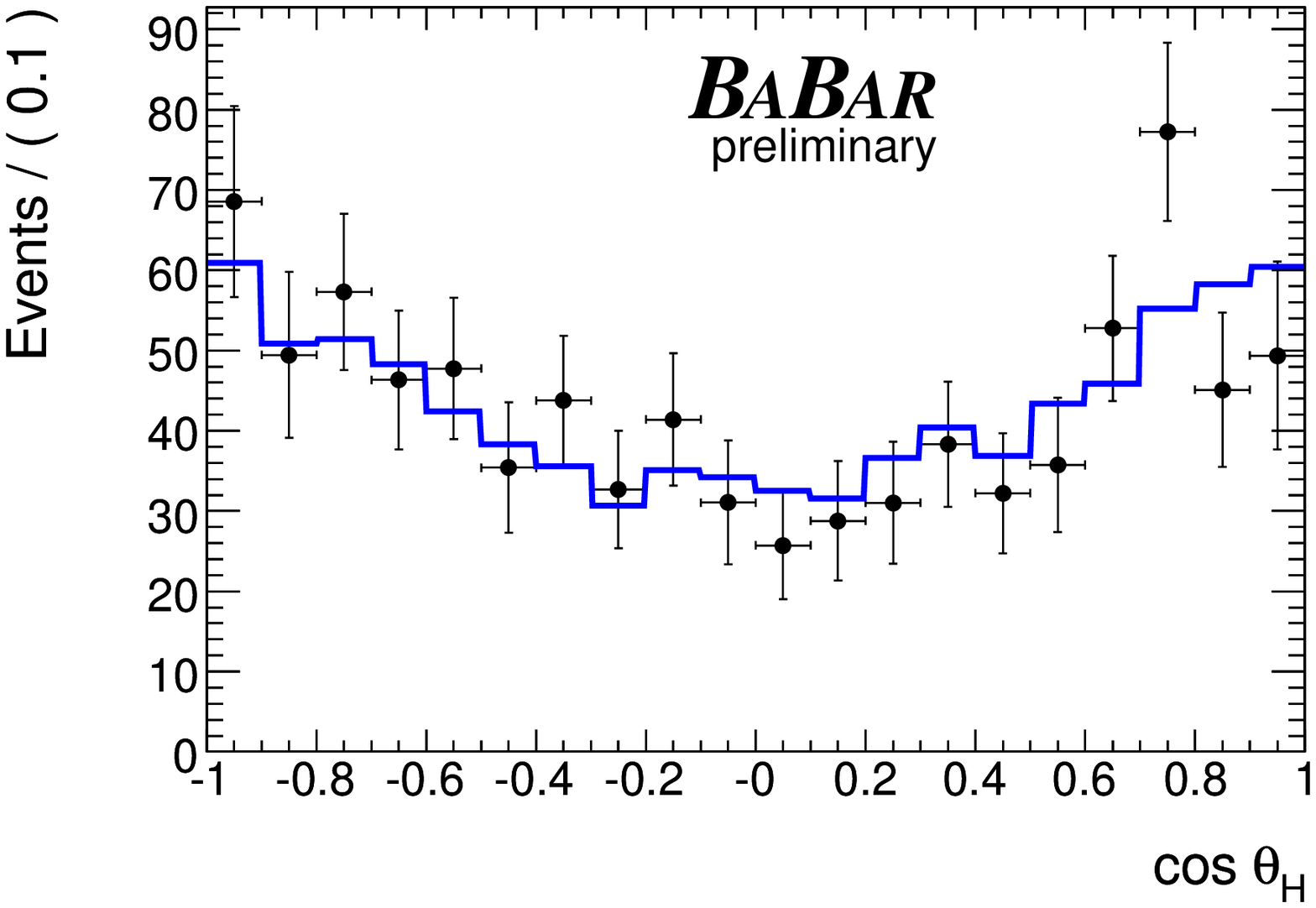} \\
    \includegraphics[width=0.23\textwidth]{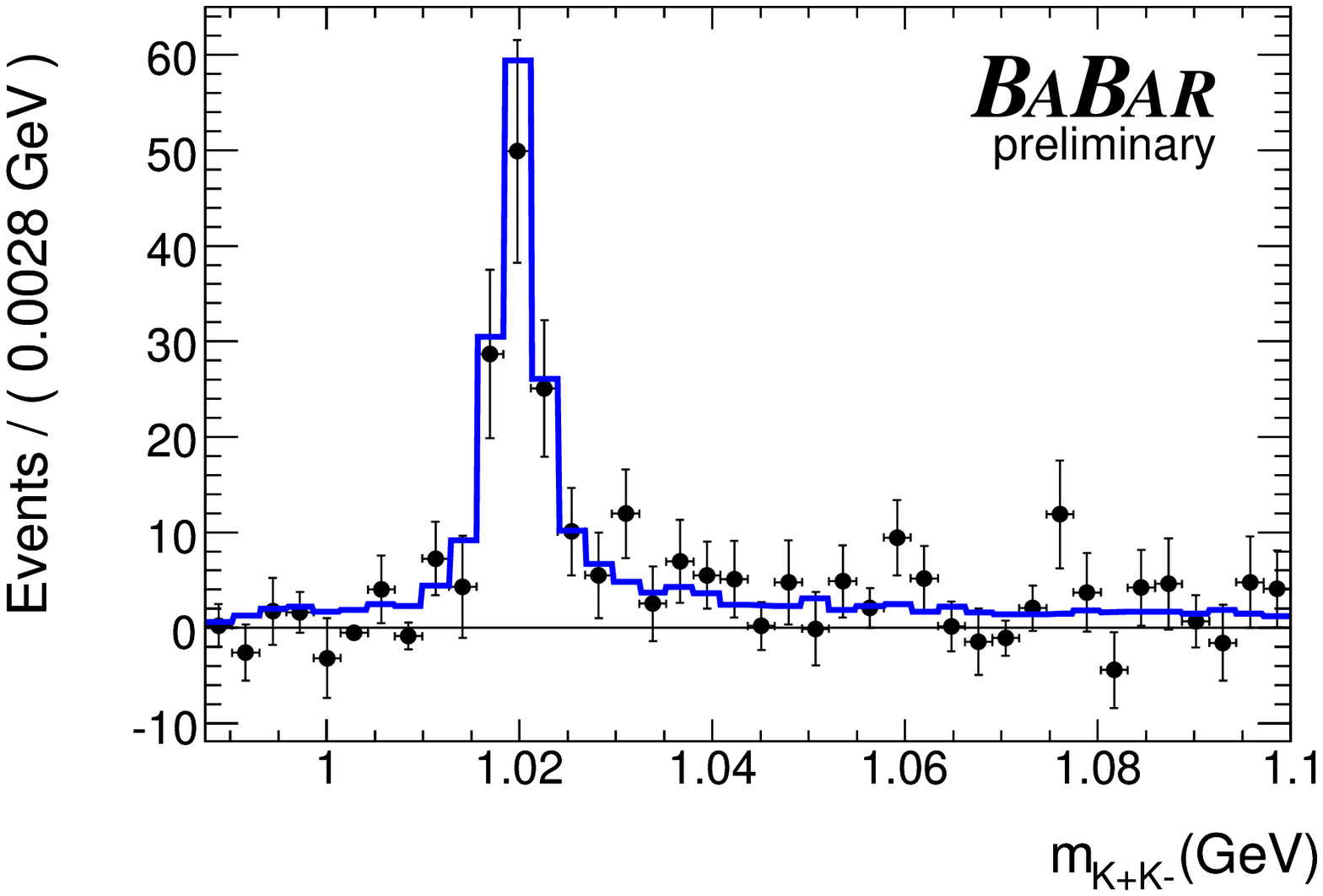} &
    \includegraphics[width=0.23\textwidth]{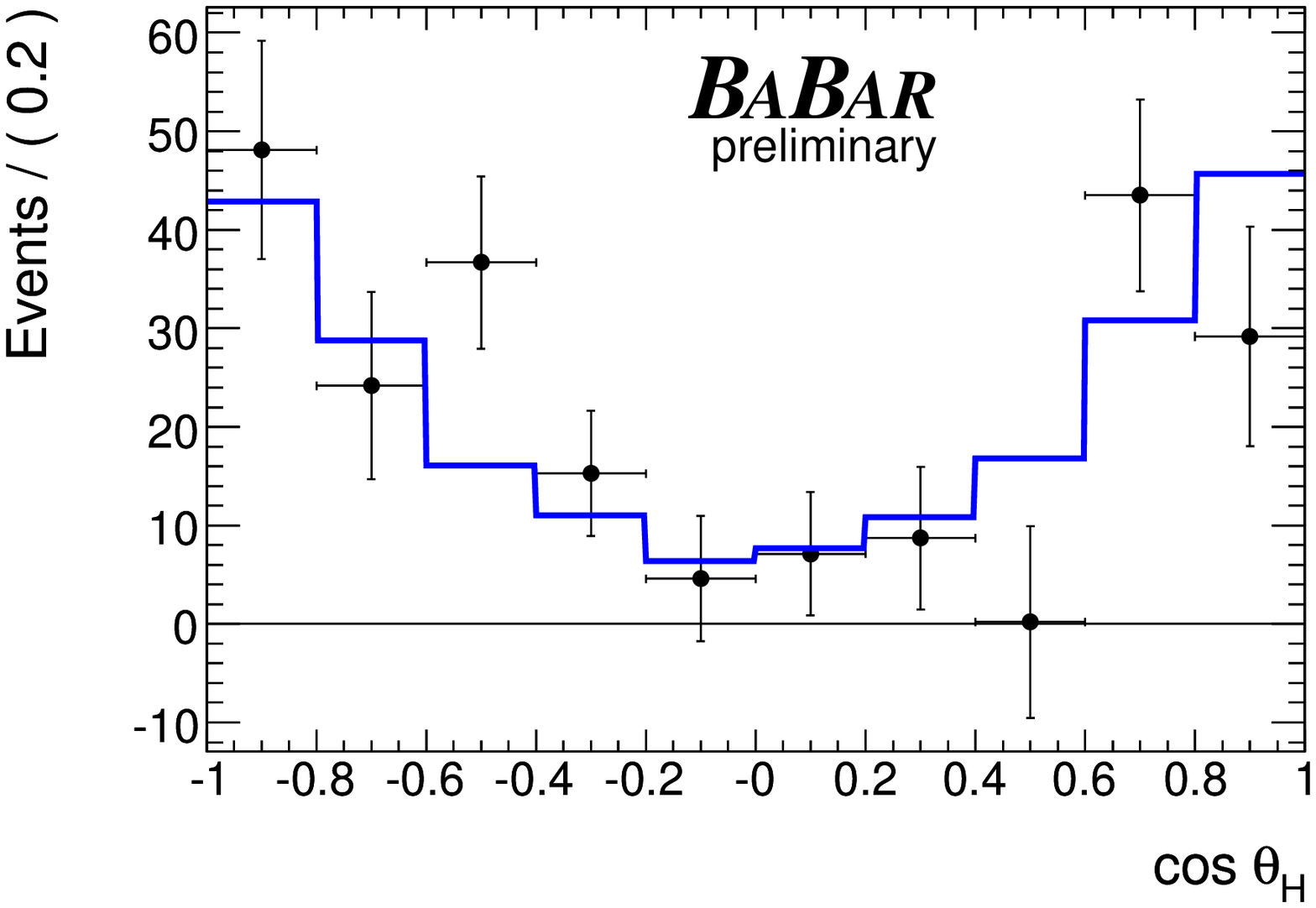} \\

  \end{tabular}
  \caption{\label{fig:DP} Distributions of the Dalitz plot variables 
    (left) \mKK and (right) \cosH for signal events (points) 
    compared with the fit PDF for \KKKspm candidates.
    Top distributions are for the entire phase space, bottom for
    a reduced region $\mKK<1.1$ \gevcc.}
\end{figure}

The fit to \babar{} data returns $879 \pm 36$ \KKKspm, $138 \pm 17$ \KKKszz 
and $499 \pm 52$ \KKKll signal candidates. The isobar amplitudes, phases and fractions
are listed in Table \ref{tab:isobars}.
Signal weighted distribution for the Dalitz plot projections 
in the entire phase space and in a reduced region $\mKK<1.1$ \gevcc, where we extract 
separate \CP asymmetry parameters, are shown in Fig. \ref{fig:DP}.

The fit to Belle data returns $840 \pm 34$ signal \KKKspm candidates, after vetoing 
the $\phi$ with the requirement $\vert \mKK-m_\phi \vert >$ 15 \mevcc.

\begin{table}
\caption{\label{tab:isobars}
Isobar amplitudes, phases, and fractions from the fit to \babar{} data. 
The fraction for non-resonant amplitude is given for the combination of the three contributions.}
\begin{ruledtabular}
\begin{tabular}{lrrrr}
\multicolumn{2}{l}{Decay }            &     Amplitude $c_r$   	&       Phase $\phi_r$   &       Fraction ${\cal F}_r$ (\%)\\
\hline  
\multicolumn{2}{l}{$\phi(1020)\Kz$}	& $  0.0098\pm 0.0016$       &       $ -0.11 \pm 0.31$   &       $12.9 \pm 1.3$         \\ 
\multicolumn{2}{l}{$f_0(980)\Kz$}	& $  0.528 \pm 0.063$         &       $ -0.33 \pm 0.26$       &       $22.3 \pm 8.9$    \\
\multicolumn{2}{l}{$X_0(1550)\Kz$}	& $  0.130 \pm 0.025$  		&       $ -0.54 \pm 0.24$   &       $4.1 \pm 1.8$      \\ 
$NR$ &$(\Kp\Km)$      			& 1 (fixed)             	&       0 (fixed)        &                              \\
        &$(\Kp\Kz)$             	& $  0.38 \pm 0.11$      	&       $  2.01  \pm 0.28$     &       $91 \pm 19$     \\
        &$(\Km\Kz)$             	& $  0.38 \pm 0.16$   	&       $ -1.19  \pm 0.37$    &                           \\
\hline
\multicolumn{2}{l}{$\chi_{c0}\Kz$}    & $  0.0343\pm 0.0067$  	&       $ 1.29 \pm 0.41$    &       $2.84 \pm 0.77$       \\
\multicolumn{2}{l}{$\Dp\Km$}          & $  1.18\pm 0.24$    	&               --       	&       $3.18 \pm 0.89$          \\
\multicolumn{2}{l}{$\Ds\Km$}          & $  0.85\pm 0.20$     	&               --          &       $1.72 \pm 0.65$        \\
\end{tabular}
\end{ruledtabular}
\end{table}

As a cross-check of the Dalitz model extracted by the fit,
we compute angular moments and extract strengths of the partial waves in \mKK 
bins using the \KKKspm sample. 
In this approach we only assume that the two lowest partial waves are present.
We verified this assumption determining that the higher angular moments 
($\langle P_{3-5} \rangle$) are consistent with zero.
In our model, the $P$-wave contribution comes from $\phi(1020)\KS$ decays
and from non-resonant events with $\Kp\KS$ and $\Km\KS$ mass dependence.
We find that the total fraction of $P$-wave in the entire Dalitz plot is 
$f_p=0.29 \pm 0.03 (stat)$.

\boldmath
\section{\CP Asymmetry}
\unboldmath

In the Belle ``quasi-two-body'' approach, the time dependent decay rate in
Eq.~\ref{eq::dalitz_plot_rate} is simplified because the interference effects are neglected:
   \begin{eqnarray}
     \label{eqn:td}
     \lefteqn{ f_{\pm}(\deltat) \; = \; \frac{e^{-|\deltat|/\tau}}{4\tau} \times }\; \\
     && \left[ \: 1 \; \pm \;
       \: S \sin{( \deltamd\deltat)} \mp C \cos{( \deltamd\deltat)} \: \: \right] \; , \nonumber
   \end{eqnarray} 
   The parameters $C$ and $S$ describe the amount of
   \CP violation in decay and in the interference between decay with and without mixing,
   respectively.
   The SM expectations for $\Bz\to\Kp\Km\KS$ are $S=-(2f_+-1)\sin 2\beta$, where
   $f_+$ is the \CP-even fraction. 
   Using isospin relations, $f_+$ has been measured on 357 \invfb data sample,
   and gives $f_+=0.93 \pm 0.09 \mbox{(stat)} \pm 0.05 \mbox{(syst)}$ \cite{bib:isospin}.
   This result is confirmed by partial wave analysis performed on \babar{} data~\cite{bib:partialwave}.

   In the \babar{} analysis, the time-dependent fit to the Dalitz plot allows to 
   extract $\betaeff$ removing the trigonometrical ambiguity $\betaeff \to \pi/2-\betaeff$.
   In this analysis the reflection is suppressed from the interference between \CP-even
   and \CP-odd decays that give rise to a $\cos(2\betaeff)$ term in Eq.\ref{eq::dalitz_plot_rate},
   in addition to the $\sin(2\betaeff)$ terms that come from the interference decays with
   and without mixing.
   In this case we measure an average $\beta$ and $\Acp$ for the full $\Kp\Km\Kz$ phase space.

   We measure also \CP-asymmetry parameters in the region $\mKK<1.1$ \gevcc, where the $\phi$
   model dependence from the rest of the Dalitz plot is highly reduced.
   In this region, we fit the \CP time-dependent asymmetries for the $\phi(1020)$ and $f_0(980)$
   components, while we fix the ones for the low-$\KpKm$ tail of the non-resonant 
   decays to the SM expectation.
   The Dalitz plot model is fixed to the one measured in the full phase space and 
   reported in Table~\ref{tab:isobars}, with the exception of the $\phi(1020)$ isobar 
   coefficients, which are fitted simultaneously with \CP-asymmetry parameters.
   In this reduced region of the phase space we find 252 $\pm$ 19, 35 $\pm$ 9, 195 $\pm$ 33
   signal events for \KKKspm, \KKKszz and \KKKll respectively.
   The results on the \CP asymmetries are shown in Table~\ref{tab:CP}.
   The sources of systematic uncertainties are briefly described below.

\begin{figure}[ptb]
  \begin{tabular}{ll}
    \includegraphics[width=0.23\textwidth]{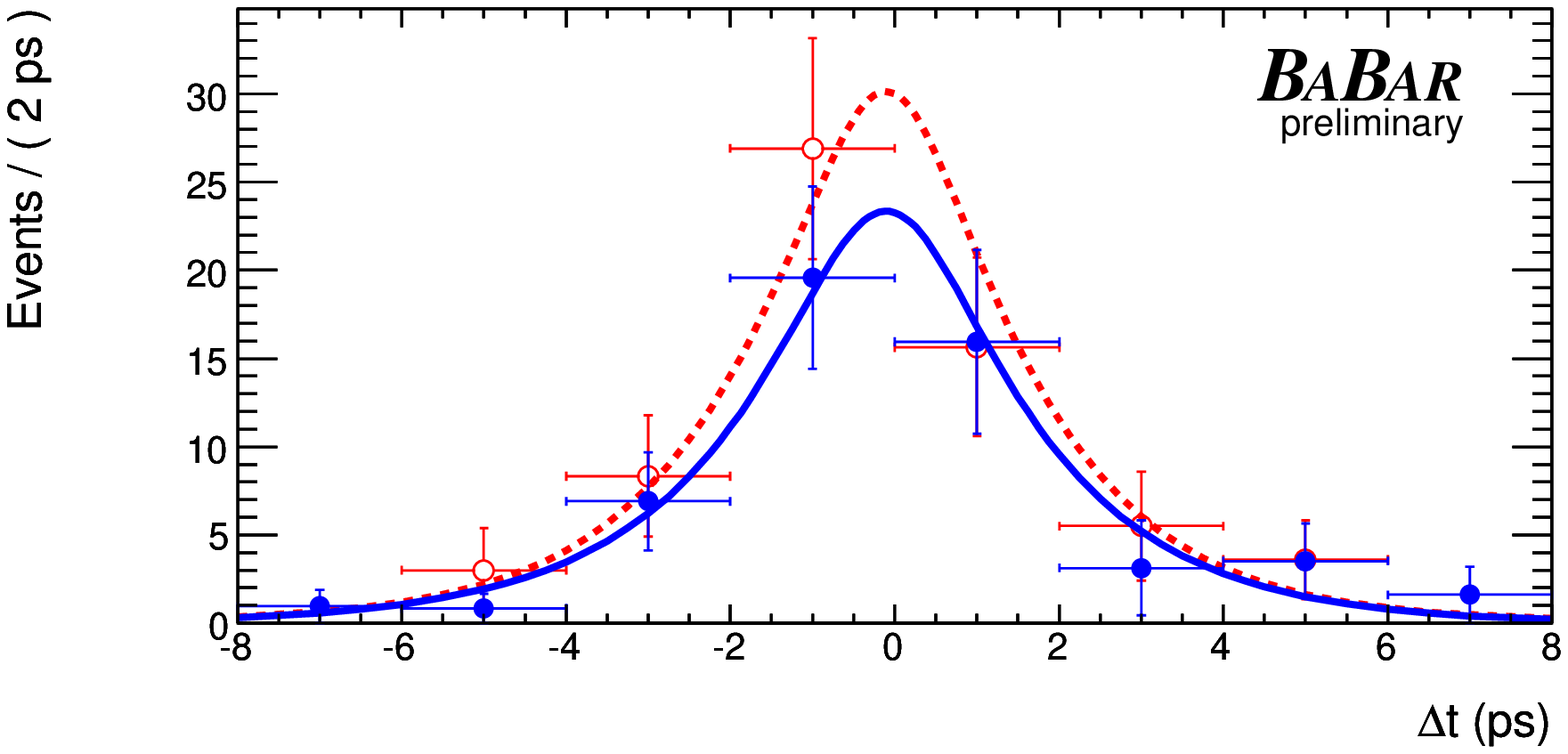}     &  
    \includegraphics[width=0.23\textwidth]{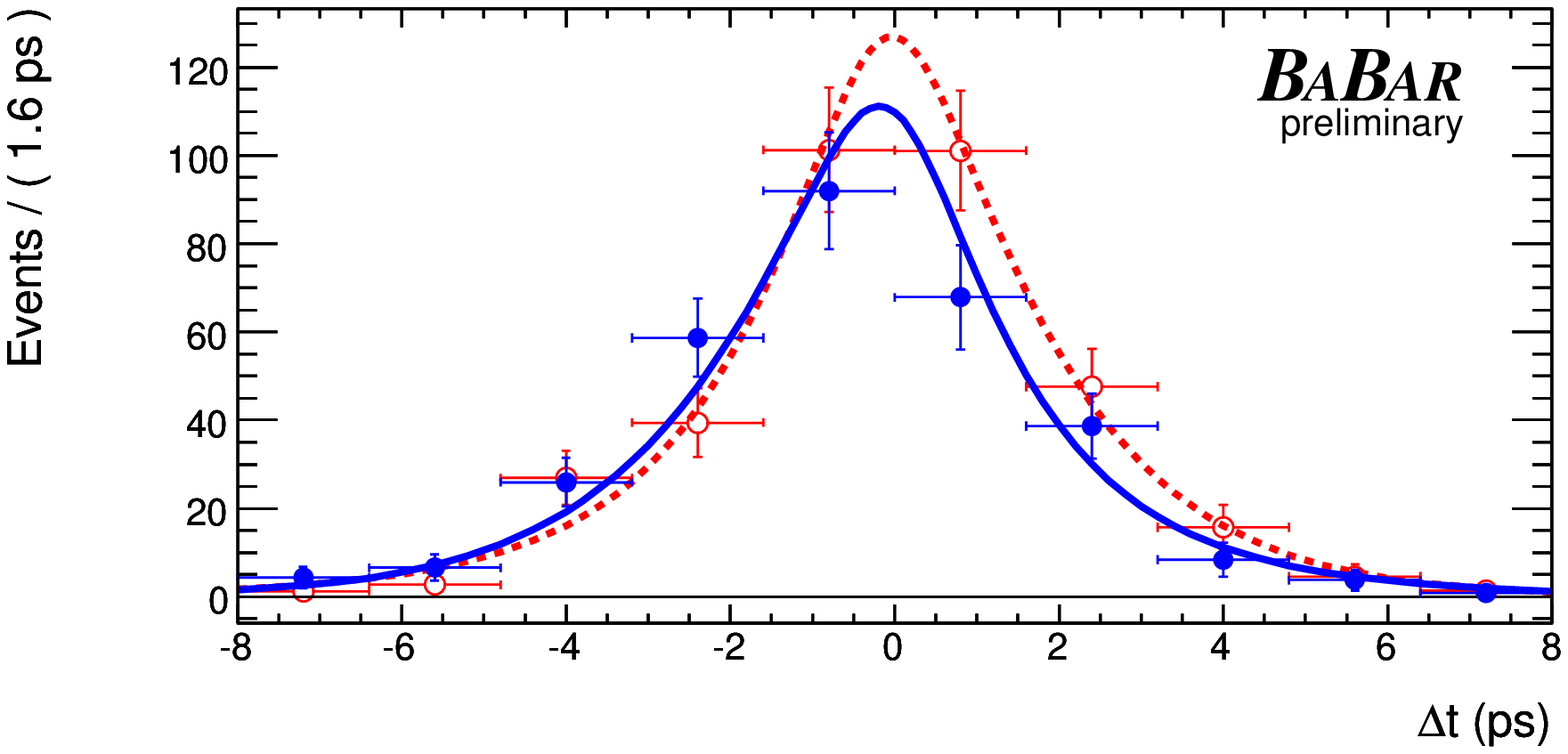}    \\
    \includegraphics[width=0.23\textwidth]{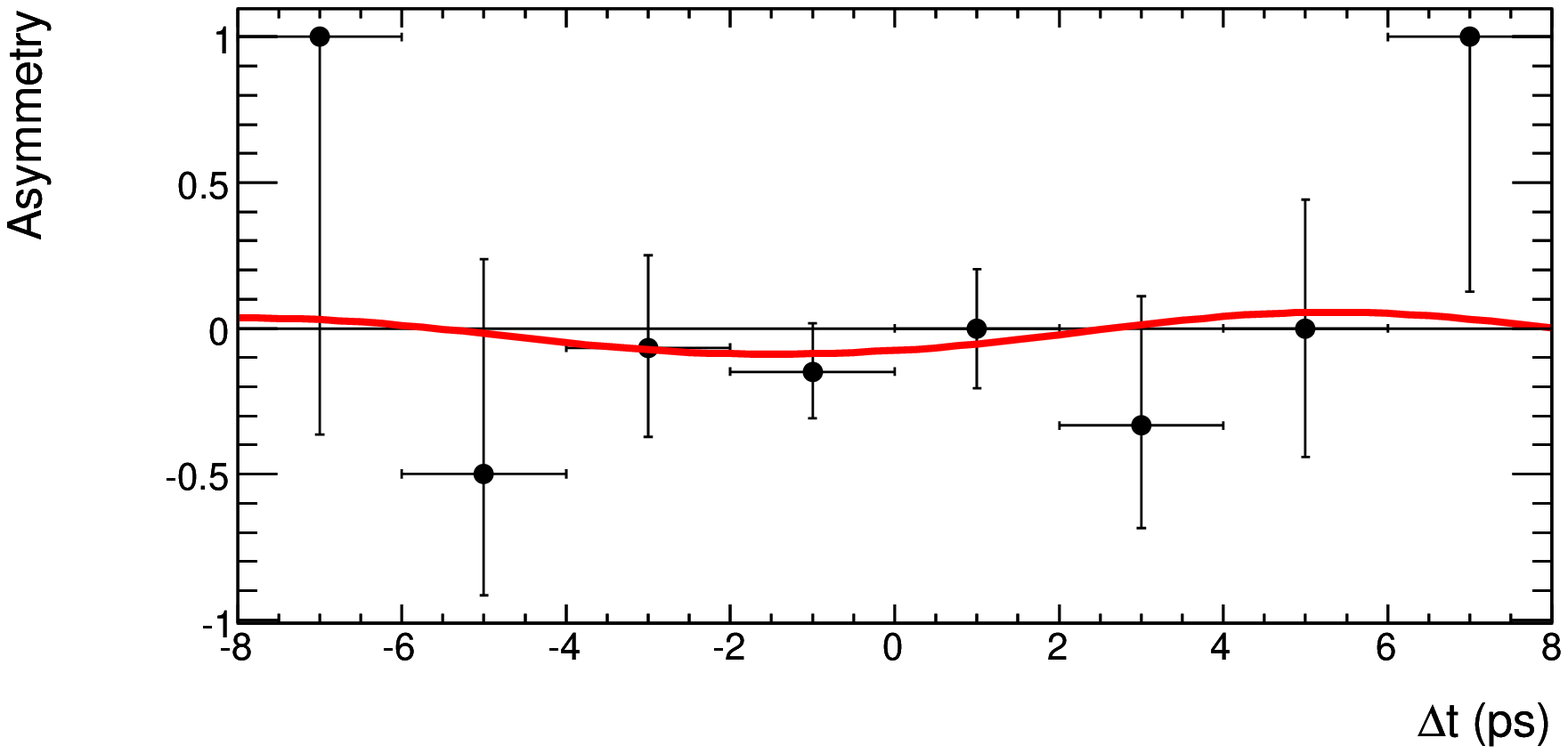} & 
    \includegraphics[width=0.23\textwidth]{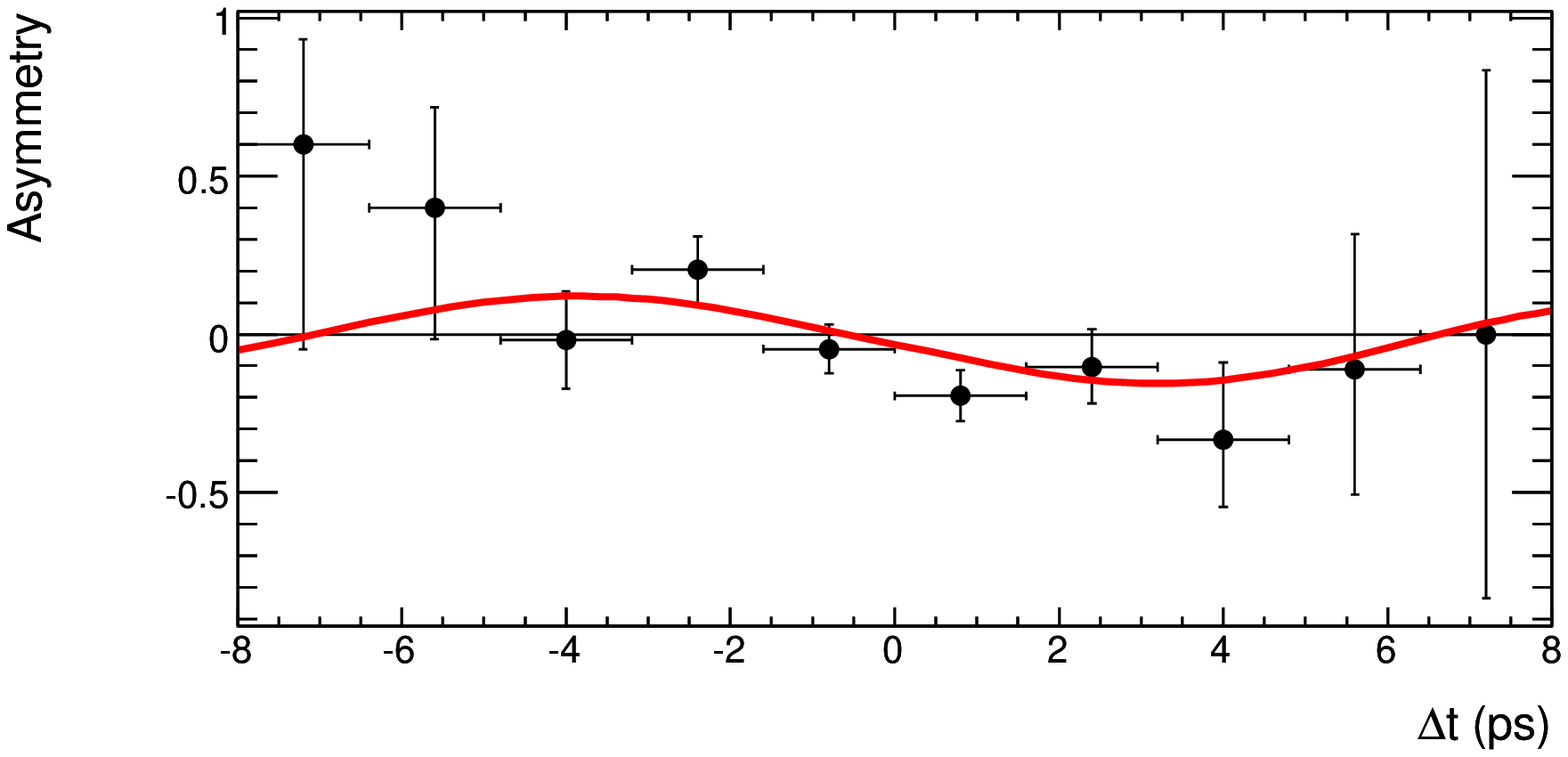}
  \end{tabular}
  \caption{
    (top) \deltat\ distributions and 
    (bottom) asymmetries for \KKKspm events for 
    (left) $1.0045 < \mKK < 1.0345~\gevcc$ and 
    (right) the whole Dalitz plot for \babar{} data. 
    For the \deltat\ distributions, \Bz- (\Bzb-) tagged signal-weighted events are shown as 
    filled (open) circles, with the PDF projection in solid blue (dashed red).  \label{fig:BaBarCP}}
\end{figure}

\begin{figure}[ptb]
  \center
  \includegraphics[width=0.23\textwidth]{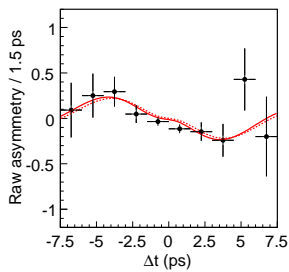}
  \caption{Asymmetry for good-tagged events ($r>0.5$) for $\Bz\to\Kp\Km\KS$ for Belle data.
  The solid curve shows the PDF projection for the result of the unbinned maximum-likelihood fit.
  The dashed curve shows the SM expectation with the measurement of \CP-violation parameters 
  for the $\Bz\to\jpsi\Kz$ decays.\label{fig:BelleCP}}
\end{figure}

   The significance of the nominal result for $\betaeff$ in the entire Dalitz plot,
   compared to the trigonometrical reflection is of 4.6$\sigma$.
   The significance of the \CP-violation is 4.5$\sigma$.
\begin{table}
\caption{\label{tab:CP}
  Time-dependent \CP-asymmetries \Acp and \betaeff for $\Bz\to\KKKz$ in $\mKK<1.1$ \gevcc 
  and in the whole phase space (\babar),
  and $C$ and $S$ for $\Bz\to\Kp\Km\KS$ in $\mKK>1.034$ \gevcc (Belle).
  The first error is statistic, the second systematic. The third error in $S$ is systematic effect due 
  to the \CP-content uncertainty.}
\begin{ruledtabular}
\begin{tabular}{cc}
Decay                           & \CP asymmetry                \\                   
\hline
$\Acp(\phi\Kz)$               &  $-0.18 \pm 0.20 \pm 0.10 $   \\
$\betaeff (\phi\Kz)$          &  $\phantom{-}0.06 \pm 0.16 \pm 0.05 $   \\
$\Acp(f_0\Kz)$                &  $\phantom{-}0.45 \pm 0.28 \pm 0.10 $   \\
$\betaeff (f_0\Kz)$           &  $\phantom{-}0.18 \pm 0.19 \pm 0.04 $   \\
\hline
$\Acp (\KKKz)$                &  $-0.034 \pm 0.079 \pm 0.025$  \\
$\betaeff (\KKKz)$            &  $\phantom{-}0.361 \pm 0.079 \pm 0.037$   \\ 
\hline
$C (\KpKm\KS)$                 &  $\phantom{-}0.09 \pm 0.10 \pm 0.05$     \\
$\sin(2\betaeff) (\Kp\Km\KS)$  &  $\phantom{-}0.68 \pm 0.15 \pm 0.03 ^{+0.21}_{-0.13}$ \\
\end{tabular}
\end{ruledtabular}
\end{table}
In Fig.~\ref{fig:BaBarCP} the distributions of $\Delta t$ for \Bz-tagged and \Bzb-tagged
events, and the asymmetry ${\cal A}(\Delta t)=(N_{\Bz}-N_{\Bzb})/(N_{\Bz}+N_{\Bzb})$, 
for background subtracted \babar{} data are shown.
In Fig.~\ref{fig:BelleCP} the asymmetries for good-tagged ($r>0.5$) $\Kp\Km\KS$ events 
in Belle data are shown.


Systematic effects are associated to parameterization of the signal PDFs,
possible fit bias, fixed \deltat\ resolution parameters, \Bz\ lifetime,
\Bz-\Bzb mixing and flavor tagging parameters.
Smaller errors due to beam-spot position uncertainty, detector alignment, and the
boost correction are based on studies done in charmonium decays.
In \babar{} analysis, an uncertainty is assigned to the resonant and 
non-resonant line-shapes. 
We try several alternative non-resonant models which omit some of the dependencies on
$\Kp\Kz$ and $\Km\Kz$ masses (see Eq.~\ref{eq:nr}).
We also study the effect of the uncertainty of the shape parameter $\alpha$ on the \CP\ parameters.
The non-resonant events contribute to the background under the $\phi$ but their shape
is determined from the high-mass region. We therefore omit the non-resonant
terms and take the difference from the reference fit as a systematic error
for $\phi\Kz$ and $f_0(980)\Kz$ \CP-asymmetries.
In Belle measurement, by far the largest systematic uncertainty on \CP
parameters is associated to the knowledge of the \CP-even fraction $f_+$.

\section{Conclusions}

We have measured the time-dependent \CP-asymmetries in $\Bz\to\KKKz$ decays,
with a simultaneous analysis of the Dalitz plot distribution of the intermediate states
(\babar) or with a ``quasi-two-body'' approach (Belle).
The measured value of \CP-asymmetries in the entire Dalitz plot is 
$\betaeff=0.361 \pm 0.079 \pm 0.037$, which is consistent with the SM expectations
($\beta \sim 0.38$). 
The trigonometrical ambiguity in \betaeff is removed at 4.6$\sigma$.
This is the first such measurement in penguin modes.
Additionally, we extracted the \CP-asymmetry parameters for $\Bz\to\phi\Kz$ and 
$\Bz\to f_0(890)\Kz$, to be $\betaeff=0.06 \pm 0.16 \pm 0.05$ and 
$\betaeff=0.18 \pm 0.19 \pm 0.04$, respectively.
Therefore we do not observe significant deviations from the SM predictions.

In Belle measurement, the measured $\sin(2\betaeff)=0.68 \pm 0.15 \pm 0.03 ^{+0.21}_{-0.13}$
for $\Bz\to\Kp\Km\KS$ decays with exclusion of $\phi\KS$ events is also consistent 
with SM predictions.

\begin{acknowledgments}

The author thanks the organizers of the workshop, and also 
Fernando Ferroni, Maurizio Pierini, Gianluca Cavoto, and Denis Dujmic, 
without that I would have not participated to it
and that let me to be involved in this fascinating measurement.
\end{acknowledgments}

\end{document}